\documentclass[twocolumn,superscriptaddress,showpacs,pra]{revtex4}
\usepackage{epsfig}
\usepackage{delarray}
\usepackage{amsmath}
\usepackage{bm}

\newcommand{\ket}[1]{| \, #1 \rangle}
\newcommand{\bra}[1]{ \langle #1 \,  |}

\begin{document}
\title{An elementary optical gate for expanding entanglement web}

\author{Toshiyuki Tashima}
\affiliation{Division of Materials Physics, Department of Materials Engineering Science,
Graduate School of Engineering Science, Osaka University, Toyonaka, Osaka 560-8531, Japan}
\affiliation{CREST Photonic Quantum Information Project, 4-1-8 Honmachi, Kawaguchi, Saitama 331-0012, Japan}
\affiliation{SORST Research Team for Interacting Carrier Electronics, 4-1-8 Honmachi, Kawaguchi, Saitama 331-0012, Japan}

\author{\c{ S}ahin Kaya \"Ozdemir}
\affiliation{Division of Materials Physics, Department of Materials Engineering Science, Graduate School of Engineering Science, Osaka University, Toyonaka, Osaka 560-8531, Japan}
\affiliation{CREST Photonic Quantum Information Project, 4-1-8 Honmachi, Kawaguchi, Saitama 331-0012, Japan}
\affiliation{SORST Research Team for Interacting Carrier Electronics, 4-1-8 Honmachi, Kawaguchi, Saitama 331-0012, Japan}

\author{\\Takashi Yamamoto}
\affiliation{Division of Materials Physics, Department of Materials Engineering Science, Graduate School of Engineering Science, Osaka University, Toyonaka, Osaka 560-8531, Japan}
\affiliation{CREST Photonic Quantum Information Project, 4-1-8 Honmachi, Kawaguchi, Saitama 331-0012, Japan}

\author{Masato Koashi}
\affiliation{Division of Materials Physics, Department of Materials Engineering Science, Graduate School of Engineering Science, Osaka University, Toyonaka, Osaka 560-8531, Japan}
\affiliation{CREST Photonic Quantum Information Project,
4-1-8 Honmachi, Kawaguchi, Saitama 331-0012, Japan}
\affiliation{SORST Research Team for Interacting Carrier Electronics,
4-1-8 Honmachi, Kawaguchi, Saitama 331-0012, Japan}

\author{Nobuyuki Imoto}
\affiliation{Division of Materials Physics, Department of Materials Engineering Science, Graduate School of Engineering Science, Osaka University, Toyonaka, Osaka 560-8531, Japan}
\affiliation{CREST Photonic Quantum Information Project,
4-1-8 Honmachi, Kawaguchi, Saitama 331-0012, Japan}
\affiliation{SORST Research Team for Interacting Carrier Electronics,
4-1-8 Honmachi, Kawaguchi, Saitama 331-0012, Japan}

\date{\today}
\begin{abstract}
We introduce an elementary optical gate
for expanding
polarization entangled W states, in which every pair of photons are
 entangled alike.
The gate is composed of a pair of 50:50 beamsplitters and
ancillary photons in the two-photon Fock state. By seeding one of
the photons in an $n$-photon W state into this gate, we obtain an
$(n+2)$-photon W state after post-selection. This gate gives a
better efficiency and a simpler implementation than previous
proposals for $\rm W$-state preparation. 
\end{abstract}
\maketitle
\section{Introduction} Quantum entanglement lies at the heart of
most of the quantum information processing tasks, e.g.,
teleportation \cite{s0}, key distribution (QKD) \cite{s1}, and
computation \cite{s2}. Entanglement between a pair of systems is
fairly simple, in the sense that there is a maximally entangled
state from which any entangled state can be generated under local
operations and classical communication (LOCC). In contrast,
multipartite entanglement among three or more systems exhibits
more variety. For example, three qubits can be entangled in two
inequivalent ways, namely, the GHZ state $\ket{\rm
GHZ}=(\ket{000}+\ket{111})/\sqrt{2}$ and the W state $\ket{\rm
W}=(\ket{001}+\ket{010}+\ket{100})/\sqrt{3}$ can never be
converted to each other under LOCC, even probabilistically
\cite{s7}.

\begin{figure}[tb]
\begin{tabular}{cc}
\begin{minipage}{0.5\hsize}
\begin{center}
\includegraphics[scale=1]{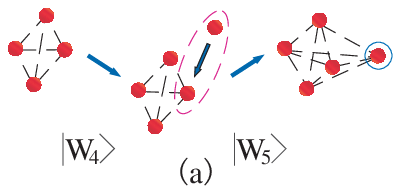}
\end{center}
\end{minipage}
\begin{minipage}{0.5\hsize}
\begin{center}
\includegraphics[scale=1]{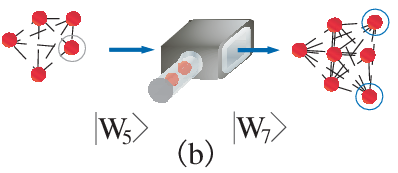}
\end{center}
\end{minipage}
\end{tabular}
\begin{tabular}{cc}
\begin{minipage}{0.5\hsize}
\begin{center}
\includegraphics[scale=1]{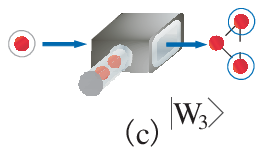}
\end{center}
\end{minipage}
\begin{minipage}{0.5\hsize}
\begin{center}
\includegraphics[scale=1]{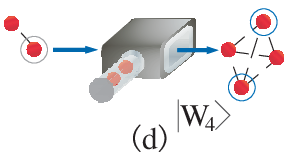}
\end{center}
\end{minipage}
\end{tabular}
\caption{
(a) Local extention of W states.
(b) The proposed optical gate (${\rm T}^{\rm W}_{+2}$)
converts $\ket{{\rm W}_N}$ to $\ket{{\rm W}_{N+2}}$.
(c) If a photon in state $\ket{1}$ is seeded, the gate produces
$\ket{{\rm W}_3}$.
(d) If we start with two photons in state
 $(\ket{01}+\ket{10})/\sqrt{2}$,
we obtain $\ket{{\rm W}_4}$.\label{fig:1s}
}
\end{figure}

The distinction between these two types of entanglement becomes
clearer if we consider their generalizations to the $N$-qubit
case: $\ket{{\rm W}_N}=\ket{N-1,1}/\sqrt{N}$ and $\ket {{\rm
GHZ}_N}=(\ket{N,0}+\ket{0,N})/\sqrt{2}$ where $\ket{N-k,k}$ is the
sum over all the terms with $N-k$ modes in $\ket{0}$ and $k$ modes
in $\ket{1}$.
In $\ket{{\rm W}_N}$, every pair of qubits are entangled with each
other directly, namely, the pairwise entanglement survives even
after the rest of the qubits are discarded \cite{s8,s9,s9.1}. In
fact, it was shown that the state $\ket{{\rm W}_N}$ has the
maximum amount of such pairwise entanglement shared by every pair
\cite{s8}. It looks as if forming a web-like structure in which
 every qubit has a bond with every other qubit
[see Fig.\ \ref{fig:1s}(a)]. On the other hand, the entanglement
in $\ket {{\rm GHZ}_N}$ is sustained by all of the $N$ qubits, and
loss of only one particle destroys the entanglement completely.
But if access to every qubit is allowed, it shows a maximal
violation of local realism \cite{s9.5}. These distinct properties
make the W and GHZ states interesting resources for multiparty
tasks and fundamental studies of quantum mechanics. Thus, there
have been many proposals and experimental implementations in
photons
\cite{s19,s20,s21,s22,s23,s24,s25,s27,s28,s29,s32,s33,s34,s35},
trapped ions \cite{s9.2,s9.4}, and NMR systems \cite{s9.3}.

The distinction also shows up when we consider how one can
increase the number of qubits forming W or GHZ states.
In the case of GHZ states,
there is a systematic way to extend its size
without accessing all of the qubits: One can pick the $N$-th qubit of
$\ket{{\rm GHZ}_N}$ and let it interact with a new qubit to produce
$\ket{{\rm GHZ}_{N+1}}$. This is not surprising since (i) the marginal
state of the remaining untouched $N-1$ qubits is the same for
$\ket{{\rm GHZ}_N}$ and $\ket{{\rm GHZ}_{N+1}}$, and (ii) the $N$-th qubit is
pivotal such that if we remove and discard it, the rest of the
qubits will be disentangled.
On the other hand, it is not so trivial
whether such a local extension of W states is
possible or not. For one thing, the marginal states of $N-1$ qubits
are different for $\ket{{\rm W}_{N}}$ and $\ket{{\rm W}_{N+1}}$. Hence no
unitary operation on the $N$-th qubit and a new qubit makes
$\ket{{\rm W}_{N+1}}$. In addition, newly added qubits must form
the pairwise entanglement with each of the uninteracted $N-1$
qubits [see Fig.\ \ref{fig:1s}(a)].

In this paper, we show that such a local extension of
polarization-entangled photonic W states is possible using a
surprisingly simple probabilistic gate composed of a two-photon
Fock state, two 50:50 beam splitters (BS), and a phase shifter
(PS), based on post-selection. Interestingly, the same gate can be
used for the expansion $\ket{{\rm W}_{N}}\rightarrow\ket{{\rm
W}_{N+2}}$ of  any size $N$ [see Fig.\ \ref{fig:1s}(b)]. The gate
also works for $N=1,2$ if we extrapolate the definition of W
states naturally: seeding $\ket{{\rm W}_1}=\ket{1}$ results in
$\ket{{\rm W}_{3}}$ [see Fig.\ \ref{fig:1s}(c)], and seeding an
EPR pair $\ket{{\rm W}_2}=(\ket{01}+\ket{10})/\sqrt{2}$ results in
$\ket{{\rm W}_{4}}$ [see Fig.\ \ref{fig:1s}(d)]. In particular,
the latter case is experimentally easier and more efficient than
any other linear optical scheme of generating $\ket{{\rm W}_{4}}$
proposed so far. Starting with (c) and applying (b) successively
$N-1$ times, we can prepare W-states with odd number of photons,
$\ket{{\rm W}_{2N+1}}$. In the same way, states with even number
of photons $\ket{{\rm W}_{2N}}$ can be prepared starting with (d).
Thus, in principle it is possible to prepare any $\ket{{\rm W}_N}$
using this gate.

\begin{figure}[tb]
\begin{center}
\includegraphics[scale=1]{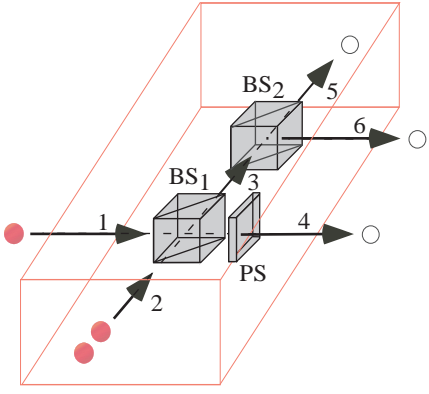}
\caption{The schematic diagram of the setup for ${\rm T}{^{\rm
W}_{+2}}$
 gate.
 \label{fig:2s}}
\end{center}
\end{figure}
\section{Working principle of the elementary gate ${T}{^{W}_{+2}}$}
In Fig.\ \ref{fig:2s}, we show the schematic of  the proposed gate.
The gate receives one photon from mode 1 as the input,
 and mixes it by a 50:50 beamsplitter (BS1) with
two ancilla photons in horizontal (H) polarization
 (we denote it by $\ket {2_{\rm H}}_2$, where
the subscript number signifies the spatial mode). One of the
output modes of BS1 is further divided into two modes by another
50:50 beamsplitter (BS2). The gate operation is successful when
each of the output modes 4, 5, and 6 has a photon. The phase
shifter (PS), which is a half-wave plate introducing a $\pi$-phase
shift between $\rm H$ and $\rm V$ (vertical) polarizations, is in
place just in order to keep the final W state in the standard
symmetric form.

 First we analyze how the gate works when the input photon
in mode 1 is H polarized ($\ket {1_{\rm H}}_1$) or V polarized
($\ket {1_{\rm V}}_1$). The action of BS1 on H polarization is
represented by the transformation $\hat{a}^{\dag}_{1{\rm
H}}=(\hat{a}^{\dag}_{3{\rm H}}-\hat{a}^{\dag}_{4{\rm H}})/\sqrt 2$
and $\hat{a}^{\dag}_{2{\rm H}}=(\hat{a}^{\dag}_{3{\rm
H}}+\hat{a}^{\dag}_{4{\rm H}})/\sqrt 2$, where
$\hat{a}^{\dag}_{j{\rm H}}$ is the photon creation operator for
mode $j$ in $\rm H$ polarization. We assume that the BS1 is
polarization-independent, namely, the transformation for $\rm V$
polarization has the same form. Using these relations, we see that
the initial states $\ket {1_{\rm H(V)}}_1\otimes\ket {2_{\rm H}}_2
=2^{-1/2}\hat{a}^{\dag}_{1{\rm H(V)}}(\hat{a}^{\dag}_{2{\rm
H}})^2\ket{0}$ evolve as
\begin{widetext}
\begin{eqnarray}
\ket {1_{\rm H}}_1\ket {2_{\rm H}}_{2}&\rightarrow&\frac {\sqrt 3}{2\sqrt2}\ket {3_{\rm H}}_3\ket {0}_4+{\underline {\frac {1}{2\sqrt2}\ket {2_{\rm H}}_3\ket {1_{\rm H}}_4}}-\frac {1}{2\sqrt2}\ket {1_{\rm H}}_3\ket {2_{\rm H}}_4-\frac {\sqrt 3}{2\sqrt2}\ket {0}_3\ket {3_{\rm H}}_4, \nonumber\\\ket {1_{\rm V}}_1\ket {2_{\rm H}}_{2}&\rightarrow&\frac {1}{2\sqrt2}\ket
{1_{\rm V}2_{\rm H}}_3\ket {0}_4+{\underline{\frac {1}{2}\ket {1_{\rm H}1_{\rm V}}_3\ket
{1_{\rm H}}_4}}+\frac {1}{2\sqrt2}\ket {1_{\rm V}}_3\ket
{2_{\rm H}}_4-{\underline{\frac {1}{2\sqrt 2}\ket {2_{\rm H}}_3\ket
{1_{\rm V}}_4}}\nonumber\\&&-\frac {1}{2}\ket {1_{\rm H}}_3\ket
{1_{\rm H}1_{\rm V}}_4-\frac{1}{2\sqrt 2}\ket
{0}_3\ket {1_{\rm V}2_{\rm H}}_4.\label{eq:1}
\end{eqnarray}
\end{widetext}
For the gate operation to be successful, there must be two photons
in mode 3 and one photon in mode 4. Hence we are interested only
in the underlined terms. The states $\ket {2_{\rm H}}_3$ and $\ket
{1_{\rm H}1_{\rm V}}_3$ appearing in the underlined terms are
transformed at $\rm BS_{2}$ as
\begin{widetext}
\begin{eqnarray}
\ket {2_{\rm H}}_3&\rightarrow&\frac {1}{2}\ket {2_{\rm H}}_5\ket {0}_6+\underline {\frac {1}{\sqrt 2}\ket {1_{\rm H}}_5\ket {1_{\rm H}}_6}+\frac {1}{2}\ket {0}_5\ket {2_{\rm H}}_6,\nonumber\\
\ket {1_{\rm H}1_{\rm V}}_3&\rightarrow&\frac {1}{2}\ket {1_{\rm H}1_{\rm V}}_5\ket
{0}_6+\underline {\frac {1}{2}\ket {1_{\rm H}}_5\ket
{1_{\rm V}}_6}+ \underline {\frac {1}{2}\ket {1_{\rm V}}_5\ket
{1_{\rm H}}_6}+\frac {1}{2}\ket {0}_5\ket
{1_{\rm H}1_{\rm V}}_6.\label{eq:1.1}
\end{eqnarray}
\end{widetext}
 Clearly, only the underlined terms in Eq.\ (\ref{eq:1.1}) contributes to
 the successful operation.
Therefore, if we postselect the successful events, the action of the gate
is given by the following state transformations:
\begin{widetext}
\begin{eqnarray}
\ket {1_{\rm H}}_1\ket {2_{\rm H}}_{2}&\rightarrow&\frac
{1}{4}\ket {1_{\rm H}}_4\ket {1_{\rm H}}_5\ket {1_{\rm H}}_6,
\label{eq:2a}\\\ket {1_{\rm V}}_1\ket {2_{\rm
H}}_{2}&\rightarrow&\frac {1}{4}\ket {1_{\rm H}}_4\ket {1_{\rm
H}}_5\ket {1_{\rm V}}_6+\frac {1}{4}\ket {1_{\rm H}}_4\ket {1_{\rm
V}}_5\ket {1_{\rm H}}_6+\frac {1}{4}\ket {1_{\rm V}}_4\ket {1_{\rm
H}}_5\ket {1_{\rm H}}_6,\label{eq:2b}
\end{eqnarray}
\end{widetext}
where we have included the effect of PS.
There are two essential features in this gate operation:
One is
the symmetrization among the input photon and the ancilla photons,
and the other is that the success probability ($1/16$) for the
$\ket{1_{\rm H}}_1$ input is one third of the probability ($3/16$) for
the $\ket{1_{\rm V}}_1$ input. In other words, all the four terms appearing
in Eq. (\ref{eq:2a}) and Eq. (\ref{eq:2b}) have the same amplitude.

The above calculation may be physically understood as follows.
When the input is a V-polarized photon, we can always determine the origin of an output photon, namely,
distinguish whether it has come from the mode 1 or the mode 2
by looking at its polarization. Hence the result is the same as in
the case of classical distinguishable particles, and the symmetrization
in Eq.~(\ref{eq:2b}) can be understood classically.
On the other hand, when the input is an H-polarized photon,
quantum interference comes
into play. In this case, there are three indistinguishable paths leading to
the final state $\ket
{1_{\rm H}}_4\ket {1_{\rm H}}_5\ket {1_{\rm H}}_6$,
depending on which one of the three photons originates
from the input mode 1. One of the path (the input photon going to
mode 4) bears the opposite sign from the other two, and
hence the probability becomes one third due to destructive
interference.
\section{Seeding and expanding polarization entangled W-states} Since
the right hand side of Eq. (\ref{eq:2b}) is $\ket{{\rm W}_3}$, we can prepare
state $\ket{{\rm W}_3}$ with a success probability of $3/16$
by applying the gate ${\rm T}{^{\rm W}_{+2}}$ on the input photon in state
$\ket{1_{\rm V}}_1$.
If we prepare an EPR pair $(\ket {1_{\rm H}}_{0}\ket {1_{\rm V}}_{1}+\ket {1_{\rm
V}}_{0}\ket {1_{\rm H}}_{1})/\sqrt{2}$ and feed the spatial mode 1 into the ${\rm T}{^{\rm W}_{+2}}$
 gate, the two terms $\ket {1_{\rm V}}_1$ and $\ket {1_{\rm H}}_1$
should evolve coherently as in
Eq. (\ref{eq:2a}) and Eq. (\ref{eq:2b}),
leading to the $\ket{{\rm W}_4}$ state:
\begin{eqnarray}
\ket {W_4} &=&\frac {1}{2}[~\ket {1_{\rm H}}_{0}\ket {1_{\rm H}}_{4}\ket {1_{\rm H}}_{5}\ket {1_{\rm V}}_{6}\nonumber\\&&+\ket {1_{\rm H}}_{0}\ket {1_{\rm H}}_{4}\ket {1_{\rm V}}_{5}\ket {1_{\rm H}}_{6}\nonumber\\&&+\ket {1_{\rm H}}_{0}\ket {1_{\rm V}}_{4}\ket {1_{\rm H}}_{5}\ket {1_{\rm H}}_{6}\nonumber\\&&+\ket {1_{\rm V}}_{0}\ket {1_{\rm H}}_{4}\ket {1_{\rm H}}_{5}\ket {1_{\rm H}}_{6}~]\label{eq:3}
\end{eqnarray}  
with success
 probability $1/8$. These values of success probability are significant
 improvements
over other linear optics-based schemes. For instance, the most
efficient schemes so far are those in \cite{s22} and \cite{s24},
respectively for $\ket{{\rm W}_4}$ and $\ket{{\rm W}_3}$ with the
corresponding success probabilities of $2/27$ and $1/9$, which are
lower than those of our proposal.

Next we discuss how this gate can be used to expand a general
W-state $\ket{{\rm W}_N}=\ket{N-1,1}/\sqrt{N}$, where $\ket{N-1,1}$ is the sum over all the terms with
$N-1$ modes in $\ket{1_{\rm H}}$ and one mode in $\ket{1_{\rm V}}$.
This state may be rewritten as
$[\ket{N-2,1}\otimes \ket{1_{\rm H}}_1+\ket{N-1,0}\otimes
\ket{1_{\rm V}}_1]/\sqrt{N}$.
If we apply the gate ${\rm T}{^{\rm W}_{+2}}$ on mode 1, we obtain
$[\ket{N-2,1}\otimes \ket{3,0}+\ket{N-1,0}\otimes
\ket{2,1}]/4\sqrt{N}=\ket{N+1,1}/4\sqrt{N}$,
implying that the gate produces $\ket{{\rm W}_{N+2}}$
with success probability $(N+2)/(16N)$,
which approaches a constant $1/16$ when $N$ becomes large.
Note that while this probability partly comes from the
inefficiency associated with linear optics schemes,
the probabilistic nature itself plays an essential role
of updating the marginal state of each of the untouched photons
from $\rho_N\equiv N^{-1}[(N-1)\ket{1_{\rm H}}\bra{1_{\rm H}}
+\ket{1_{\rm V}}\bra{1_{\rm V}}]$ to that of $\rho_{N+2}$. By cascading ${\rm
T}{^{\rm W}_{+2}}$ gates, we can prepare
W states over 5 or more photons. Starting
with $\ket{1_{\rm V}}$ as an input
and cascading the gate $k$ times, one can prepare $(2k+1)$-photon
W state, $\ket{{\rm W}_{2k+1}}$, provided that
coincidence detection is observed at $2k+1$ output spatial modes.
The success probability of such an event is given by $p_{{\rm
success}}=(2k+1)2^{-4k}$.
Similarly, starting with an EPR pair
and cascading $k$ gates,
one can prepare $2(k+1)$-photon
W state, $\ket{{\rm W}_{2(k+1)}}$, with a success probability
of $p_{{\rm success}}=(k+1)2^{-4k}$.
Besides our current proposal,
the scheme based on $N\times N$ multiport interferometers
\cite{s24,s25} is so far the only proposal encompassing
generation of $\ket{{\rm W}_N}$ with arbitrary $N$.
 This scheme
requires a different multiport device for each $N$.
In addition, numerical calculation up to $N=7$ shows that
our proposal has better efficiency, e.g., for $N=5$ our
proposal succeeds with a probability 12 times higher than that of
the multiport interferometer. Note also that $N\times
N$ interferometer cannot generate the $\ket{{\rm W}_6}$ state because of
the zero probability of coincidence detection due to destructive
interference.

\section{Feasibility analysis for $\ket{{\rm W}_4}$}So far,
several linear optical schemes for preparing $\ket{{\rm W}_4}$
have been proposed \cite{s22,s23,s24}, but no experiments have
been done yet. It is thus interesting to consider the feasibility
of our scheme with practical photon sources, namely, parametric
down-conversion (PDC) and/or weak coherent pulses (WCP) obtained
by attenuating laser pulses. First, suppose that an EPR pair and
ancillary two photons are both generated from PDC with rates
$\gamma$ and $g$, respectively, which are $\sim 10^{-4}$ in
typical experiments. In this case,  the errors are mainly caused
by generation of three pairs of photons in total. Such events
occur with rate $O(\gamma^2 g)$ or $O(\gamma g^2)$, which is small
compared with the rate $O(\gamma g)$ of the desired
events. Alternatively, we may
also use WCP instead of PDC for the ancillary photons in mode 2.
If the mean photon number of WCP is $\nu$, the desired events
occur with rate $O(\gamma\nu^2)$. With the requirement $\nu\ll 1$ as usual, the main source of errors in
this case is two-pair production at PDC, resulting in two photons
in the input mode 1. Then, one photon in the WCP leads to triple
coincidence at modes 4, 5, and 6, which occurs with rate
$O(\gamma^2\nu)$. Thus we need to satisfy $\gamma \ll\nu\ll 1$ to
obtain a high fidelity. In both cases, the contribution of the
dark counts of detectors are negligibly small in
 current experiments \cite{s39,s40}.
Mode
mismatch effects may be minimized by proper spectral and
spatial filtering as discussed in Ref. \cite{s41}.

\section{Extending polarization entangled GHZ states}In our proposed gate, a W state is essentially produced before
the PS in Fig.~\ref{fig:2s}, which merely applies a local unitary
operation. We then notice that the essential passive components
(the two BSs) are polarization-independent, and the
polarization dependence of the gate stems solely from
the polarization of the ancilla photons, i.e., the expansion of a W state is achieved even if we rotate the
polarization of the photons in input modes 1 and 2 by the same angle. Interestingly, this indicates
the possibility of expanding states other than W state by changing the
polarization of ancillas. Indeed, the same set of the two BSs can also be used for
the extension of the GHZ states, just by replacing the ancillary state
$\ket{2_{\rm H}}_2$ with $\ket {\rm 1_{H}1_{V}}_2$.
If the photon in mode 1 is $V$-polarized, destructive two-photon
interference kills the events with only one $V$-polarized photon in mode
4. Hence, under the condition that each of the output modes 4, 5, and 6
has a photon, the transformation is given by
$\ket {1_{\rm V}}_1\to 2^{-3/2}\ket {1_{\rm V}}_4
\ket {1_{\rm V}}_5\ket {1_{\rm H}}_6$. Similarly, an $H$-polarizaed
input will be transformed as
$\ket {1_{\rm H}}_1\to 2^{-3/2}\ket {1_{\rm H}}_4
\ket {1_{\rm H}}_5\ket {1_{\rm V}}_6$.
It is then obvious that this gate achieves
$\ket{{\rm GHZ}_N}\to \ket{{\rm GHZ}_{N+2}}$ up to a local unitary,
with a success probability of 1/8.
\section{Conclusion}In summary, we have proposed a simple elementary optical gate based on
post-selection for expanding the symmetrically shared
entanglement in polarization entangled W states.
With a proper seeding, the gate can also be used for
preparation of W states,
and it has a larger success probability
than other preparation methods.
We believe that the proposed
gate is easy to implement and feasible with the current
experimental technologies.
In our gate, polarization-dependent components play no
essential role, and the desired transformation is achieved
by multi-photon interference between the input photon and
the ancilla photons. Note that this does not require sub-wavelength adjustments.We have also shown that just by
changing the state of the ancilla photons, we obtain a
gate for extending GHZ states.
\section*{Acknowledgments}
This work was supported by 21st Century COE Program by the JSPS
and by a MEXT Grant-in-Aid for Young Scientists (B) 17740265.


\begin{thebibliography}{99}
\bibitem{s0}
C. H. Bennett {\it et al}, \prl \textbf{70,} 1895 (1993).

\bibitem{s1}
A. K. Ekert, \prl \textbf{67,} 661 (1991).

\bibitem{s2}
M. A. Nielsen and I. L. Chuang, ``Quantum Computation and Quantum
    Information'', Cambridge university Press, (2000).



\bibitem{s7}
W. D\"ur, G. Vidal, and J. I. Cirac, \pra \textbf{62,} 062314 (2000).

\bibitem{s8}
M. Koashi, V. Bu$\check{\rm z}$ek, and N. Imoto, \pra \textbf{62,}
    050302(R) (2000).
\bibitem{s9}
W. D\"ur, \pra \textbf{63,}
    020303(R) (2001).
\bibitem{s9.1}
Yu-xi Liu, A. Miranowicz, M. Koashi, and N. Imoto, \pra \textbf{66,}
    062309 (2002).
\bibitem{s9.5}
N. D. Mermin, \prl \textbf{65,}
    1838 (1990).




\bibitem{s19}
T. Yamamoto, K. Tamaki, M. Koashi, and N. Imoto, \pra \textbf{66,} 064301
    (2002). The demonstration is shown in N. Kiesel {\it et al}, \prl
   \textbf{98,} 063604 (2007).
\bibitem{s20}
A. Zeilinger, M. A. Horne, H. Weinfurter, and M. Zukowski, \prl \textbf{78,} 3031
   (1997).

\bibitem{s21}
J. G. Rarity, and P. R. Tapster, \pra \textbf{59,} R35
    (1999).

\bibitem{s22}
X. Zou, K. Pahlke, and W. Mathis, \pra \textbf{66,} 044302
    (2002).

\bibitem{s23}
Y. Li, and T. Kobayashi, \pra \textbf{70,} 014301
    (2004).

\bibitem{s24}
B. -S. Shi, and A. Tomita, \jmo \textbf{52,} 755 (2005).
\bibitem{s25}
Y. L. Lim and A. Beige, \pra \textbf{71,} 062311
    (2005).


\bibitem{s27}
P. Walther, M. Aspelmeyer, and A. Zeilinger, \pra \textbf{75,} 012313
    (2007).

\bibitem{s28}
D. Bouwmeester {\it et al},  \prl\textbf{82,} 1345 (1999).

\bibitem{s29}
K. J. Resch, P. Walther, and A. Zeilinger, \prl
    \textbf{94,} 070402 (2005).

\bibitem{s32}
N. Kiesel {\it et al}, \jmo \textbf{50,} 1131 (2003).

\bibitem{s33}
M. Eibl {\it et al}, \prl  \textbf{92,} 077901 (2004).

\bibitem{s34}
H. Mikami, Y. Li, K. Fukuoka, and T. Kobayashi, \prl
    \textbf{95,} 150404 (2005).

\bibitem{s35}
K. J. Resch, P. Walther, and A. Zeilinger, \prl  \textbf{94,} 240501
    (2005).
\bibitem{s9.2}
H. H\"affner {\it et al}, Nature (London) {\bf 438},
    643 (2005).
\bibitem{s9.4}
D. Leibfried {\it et al}, Nature (London) {\bf 438},
    639 (2005).
\bibitem{s9.3}
G. Teklemariam {\it et al}, \pra \textbf{66,}
    012309 (2002).


\bibitem{s39}
\c{S}. K. \"Ozdemir, A. Miranowicz, M. Koashi, and N. Imoto, \pra
    \textbf{64,} 063818 (2001).

\bibitem{s40}
T. Yamamoto, M. Koashi, and N. Imoto, \pra  \textbf{64,} 012304
(2001).

\bibitem{s41}
\c{ S}. K. \"Ozdemir, A. Miranowicz, M. Koashi, and N. Imoto, \pra
\textbf{66,} 053809 (2002).
\end{thebibliography}
\end{document}